\documentclass[12pt]{article}
\newcommand{\be}{\begin{equation}}
\newcommand{\ee}{\end{equation}}
\newcommand{\bea}{\begin{eqnarray}}
\newcommand{\eea}{\end{eqnarray}}
\newcommand{\sptwo}{1.4}
\newcommand{\doublespace}{\edef\baselinestretch{\sptwo}\Large\normalsize}
\begin{document}
\begin{center}
{\large\bf The Effective Action\\
For\\
Brane Localized Gauge Fields}
\end{center}
~\\
\begin{center}
{\bf T.E. Clark\footnote{e-mail address: clark@physics.purdue.edu}} 
and
{\bf Muneto Nitta\footnote{e-mail address: nitta@physics.purdue.edu}}\\
{\it Department of Physics\\
Purdue University\\
West Lafayette, IN 47907-1396}\\
~\\
And\\
~\\
{\bf T. ter Veldhuis\footnote{e-mail address: veldhuis@physics.umn.edu}}\\
{\it Department of Physics\\
University of Minnesota\\
Minneapolis, MN 55455}
~\\
~\\
\end{center}
\begin{center}
{\bf Abstract}
\end{center}
The low energy effective action including gauge field degrees of freedom
on a non-BPS p=2 brane embedded in a N=1, D=4 target superspace is obtained
through the method of nonlinear realizations of the associated super-Poincar\'{e}
symmetries.  The invariant interactions of the gauge fields and the brane
excitation modes corresponding to the Nambu-Goldstone degrees of freedom
resulting from the broken space translational symmetry and the target
space supersymmetries are determined.  Brane localized matter field interactions
with the gauge fields are obtained through the construction of the combined
gauge and super-Poincar\'e covariant derivatives for the matter fields.
~\\

\newpage
\doublespace

\section*{}

When a target manifold contains an embedded defect its symmetry group is
spontaneously broken to those of the world volume of the defect and its 
complement. The low energy quantum fluctuations of the defect in directions 
associated with broken symmetry generators
correspond to Nambu-Goldstone degrees of freedom. It is well
known that defects, such as an embedded domain wall, may cause
the localization of additional massless as well as massive
scalar and fermionic degrees of freedom on their world volume
\cite{Polyakov:ek,Voloshin:aa,Jackiw:1975fn}.
The idea that such localized degrees of freedom constitute the
matter of our universe is explored in brane-world scenarios
\cite{Rubakov:bb,Akama:jy}.
While in a field theoretical context defects embedded in a flat (non-compact)
target space do not yield localized
gauge fields as zero modes \cite{Dubovsky:2001pe},
several alternative mechanisms that result in (quasi-) localized
non-Abelian gauge fields on the defect have
been proposed \cite{Dvali:1996xe,Dvali:2000rx,Akhmedov:2001ny}.
(Kaluza-Klein localization of the gauge field zero mode for compact or finite 
volume spaces is readily realized \cite{Oda:2000zc}.) In any realistic 
brane-world scenario localized gauge fields are a physical necessity, and, 
regardless of the underlying physics, gauge fields are required to be present 
in the effective world volume field theory. The invariance of the world volume 
action under target space and gauge symmetries then dictates the form of the 
interaction between the gauge fields and the Nambu-Goldstone fields.

The purpose of this brief report is
to extend the construction of the low energy effective action of a non-BPS
p=2 brane embedded in a N=1, D=4 target superspace of reference
\cite{Clark:2002bh} to include gauge fields in addition to the 
Nambu-Goldstone and matter degrees of freedom.
This is done in a model independent way through the method of non-linear
realizations of the spontaneously broken target space super-Poincar\'e
symmetries. 
Actions for non-BPS D-branes have been constructed in 
reference \cite{Sen:1999md}.  There the Green-Schwarz 
action \cite{Green:1983wt}, excluding the Wess-Zumino term, was utilized 
to find the action for a Dp-brane embedded in D=10 superspace.  The relation 
of the generalized Green-Schwarz method of \cite{Sen:1999md} to the coset 
method employed here, and in reference \cite{Clark:2002bh}, remains to 
be elucidated.  However, in the BPS superparticle case, the relation between the 
Green-Schwarz action and that obtained using nonlinear realization techniques 
has been clarified in references \cite{Ivanov:1999fw}.  
Returning to the non-BPS p=2 brane at hand, the 
Poincar\'e symmetries of the D=3 world volume space-
time of the brane, which has static gauge coordinates $x^m$, $m=0,1,2$,
are realized linearly. The broken translation symmetry Nambu-Goldstone 
boson is denoted by $\phi (x)$ and the
broken SUSY Goldstino D=3 (real) Majorana fields are denoted by $\theta_i (x)$ and
$\lambda_i(x)$, $i=1,2$.  In addition there is an auxiliary Nambu-Goldstone D=3
vector field, $v^m (x)$, associated with the broken D=4 Lorentz transformations.
It can be expressed in terms of the independent degrees of freedom $\phi$,
$\theta$ and $\lambda$ through the \lq\lq inverse Higgs mechanism"
\cite{Ivanov:1975zq}.

The non-linear realization of the N=1, D=4 super-Poincar\'e symmetry group $G$, 
with its broken space translation symmetry and the broken
supersymmetries, induces a Nambu-Goldstone field dependent world volume
general
coordinate transformation
\be
dx^{\prime m} = dx^n G_n^{~m} ,
\ee
where \cite{Clark:2002bh}
\be
G_n^{~m} = \frac{\partial x^{\prime m}
}{\partial x^n} = \delta_n^{~m} - i (\partial_n \theta \gamma^0 \gamma^m \xi +
\partial_n
\lambda \gamma^0 \gamma^m \eta ) -\partial_n \phi b^m
+ \epsilon_n^{~~ms} \alpha_s .
\ee
Concurrent with this are the non-linear transformations of the
Nambu-Goldstone fields
\bea
\Delta \phi &=& z + (\xi \gamma^0 \lambda - \theta \gamma^0 \eta ) - b^m
x_m
\cr
\Delta\theta_i &=& \xi_i - \frac{i}{2} b_m (\gamma^m \lambda )_i - i\rho
\lambda_i  -\frac{i}{2} \alpha_m (\gamma^m \theta )_i \cr
\Delta\lambda_i &=& \eta_i + \frac{i}{2} b_m (\gamma^m \theta )_i + i\rho
\theta_i  -\frac{i}{2} \alpha_m (\gamma^m \lambda )_i \cr
\Delta v^m &=& +\frac{\sqrt{v^2}}{\tanh{\sqrt{v^2}}} \left( b^m -
\frac{v^r b_r v^m}{v^2} \right) + \frac{v^r b_r v^m}{v^2} + \epsilon^{mnr}
\alpha_n v_r .
\eea
where the set of transformation parameters $\{z, a^m , (\xi_i , \eta_i ),
b^m ,\alpha^m , \rho \}$ correspond to the
\{broken D=4 translation, unbroken D=3 space-time translations, (broken SUSY,
broken SUSY), broken D=4 Lorentz rotations,
unbroken D=3 Lorentz rotations, unbroken $R$\} transformations of the N=1, D=4
super-Poincar\'e symmetries.  (Since the unbroken symmetries are those of
the D=3 Poincar\'e group, the D=4 transformation parameters and the
Nambu-Goldstone fields are expressed in terms of their D=3 Lorentz group 
transformation properties.) The intrinsic variation of the fields, 
$\delta \varphi \equiv \varphi^\prime (x) -\varphi (x)$, is related to the 
above total variation, $\Delta \varphi = \varphi^\prime (x^\prime) - \varphi (x)$, 
by the Taylor expansion shift in the space-time
coordinates: $\delta \varphi = \Delta \varphi - \delta x^m \partial_m
\varphi$, with $\delta x^m = x^{\prime m} - x^m 
= a^m - i ( \bar{\xi} \gamma^m \theta + \bar{\eta} \gamma^m \lambda)
- \phi b^m + \epsilon^{mnr} \alpha_n x_r$.

According to the coset construction method \cite{Coleman:sm,Volkov73},
the dreibein, the covariant derivatives of the Nambu-Goldstone fields 
and the spin connection can be obtained from the Maurer-Cartan one-forms.  
The covariant world volume coordinate differentials, $\omega^a$,
are found in terms of the dreibein, $e_m^{~a}$, and the world-volume
coordinate differentials,
$dx^m$,
\be
\omega^a = dx^m e_m^{~a} ,
\ee
where the dreibein has the factorized form $e_m^{~a} =
\hat{e}_m^{~b} N_b^{~a}$ \cite{Clark:2002bh}.  The Akulov-Volkov dreibein
$\hat{e}_m^{~a}$
is
\be
\hat{e}_m^{~a} =A_m^{~a}= \delta_m^{~a} + i \partial_m \theta
\gamma^0\gamma^a \theta + i \partial_m \lambda \gamma^0 \gamma^a \lambda ,
\ee
while the Nambu-Goto dreibein $N_a^{~b}$ has the form
\be
N_a^{~b} = \delta_a^{~b} +\left[ \cosh{\sqrt{v^2}} -1\right] \frac{v_a
v^b}{v^2} + \left( {\hat{\cal D}}_a \phi +   {\hat{\cal D}}_a \theta
\gamma^0 \lambda - \theta \gamma^0 {\hat{\cal D}}_a \lambda \right)  v^b
\frac{\sinh{\sqrt{v^2}}}{\sqrt{v^2}} ,
\label{N}
\ee
where ${\hat{\cal D}}_a = \hat{e}_a^{-1 m}\partial_m = A_a^{-1 m}
\partial_m$ is the Akulov-Volkov partial (SO(1,2)) covariant derivative.  
In addition the Maurer-Cartan one-form defines the covariant derivatives 
of the Nambu-Goldstone fields. In particular, 
the fully (SO(1,3)) covariant derivative, 
$\nabla_a \phi$, of the translational Nambu-Goldstone boson $\phi$ can be 
expanded in terms of the Akulov-Volkov partially covariant coordinate 
differential basis one-forms, $\hat{\omega}^a = dx^m \hat{e}_m^{~a} $,
\be
N_a^{~b} \nabla_b \phi = \cosh{\sqrt{v^2}} \left[ \left( \hat{\cal D}_a
\phi
+ \hat{\cal D}_a \theta \gamma^0 \lambda -\theta \gamma^0 \hat{\cal D}_a
\lambda \right)   + v_a \frac{\tanh{\sqrt{v^2}}}{\sqrt{v^2}}\right] .
\ee
Setting $\nabla_a \phi = 0$ results in the \lq\lq inverse Higgs mechanism"
\cite{Ivanov:1975zq}, allowing $v^m$ to be expressed in terms of the
independent Nambu-Goldstone degrees of freedom, $\phi$, $\theta$ and $\lambda$, as
\be
v_a \frac{\tanh{\sqrt{v^2}}}{\sqrt{v^2}} = -\left( \hat{\cal D}_a \phi +
\hat{\cal D}_a \theta \gamma^0 \lambda -\theta \gamma^0 \hat{\cal D}_a
\lambda \right)  .
\label{inversehiggs}
\ee

The covariant world volume coordinate differentials $\omega^a$ transform
under the N=1, D=4 super-Poincar\'e group $G$ as structure group vectors
\be
\omega^{\prime b} = \omega^a L_a^{~b} ,
\label{structuretrans}
\ee
where
\be
L_a^{~b} = (e^{-i\beta^c \tilde{M}_c})_a^{~b} ,
\label{L}
\ee
with the D=3 Lorentz vector representation matrix $(\tilde{M}_{c})_a^{~~b}
= i \epsilon_{ca}^{~~~b}$.  The field dependent transformation parameter
$\beta^c$ is  determined from the non-linear realization of $G$ on the coset
element $\Omega \in G/H$ through $g\Omega = \Omega^\prime h$, with the group 
element $g\in G$ corresponding to the transformation, $\Omega^\prime \in G/H$
the transformed coset element, 
and $h\in H=SO(1,2)\otimes R$, the unbroken D=3 Lorentz and $U_R(1)$ symmetry
groups.  For infinitesimal D=4 Lorentz transformations, with D=3 parameters
$\alpha^m$ for the unbroken transformations and $b^m$ for the broken ones, 
the induced local Lorentz transformation parameter was determined in 
\cite{Clark:2002bh} to be
\be
\beta^a (g,v)= \alpha^a - \frac{1}{2}
\frac{\tanh{\frac{1}{2}\sqrt{v^2}}}{\frac{1}{2}\sqrt{v^2}} \epsilon^{abc}
b_b
v_c .
\label{beta}
\ee

Correspondingly, under a $G$-transformation the dreibein, $e_m^{~a}$,
transforms with one world index and one tangent space (structure group) index as
\be
e_m^{\prime ~a}= G_m^{-1 n} e_n^{~b} L_b^{~a} ,
\ee
and likewise for the inverse dreibein
\be
e^{\prime -1 m}_a = L_a^{-1 b} e_b^{-1 n} G_n^{~ m} .
\ee
Utilizing the flat tangent space metric, $\eta_{ab}$, the induced world
volume metric tensor is given in terms of the dreibein as
\be
g_{mn} = e_m^{~a} \eta_{ab} e_n^{~b} .
\ee
Consequently the invariant interval can be expressed as
$ds^2 = dx^m g_{mn} dx^n$.
The leading term in the D=4 super-Poincar\'e invariant action is given by
integrating the constant brane tension, $\sigma$, over the area of the brane
\be
\Gamma = -\sigma \int d^3 x \det{e}.
\ee

The matter fields localized on the brane are characterized by their D=3
Lorentz group (with generators $M^a$) transformation properties.  A scalar
field, $S(x)$, is in the trivial representation of the Lorentz group: 
$M^a \rightarrow (\tilde{M}^a)=0$. Fermion fields, $\psi_i (x)$, are in the 
spinor representation: $M^a \rightarrow (\tilde{M}^a)_{ij}=-1/2 \gamma^a_{ij}$.
Each matter field, $M(x)$, transforms under $G$ as
\be
M^\prime (x^\prime ) \equiv \tilde{h} M(x) ,
\ee
where $\tilde{h}$ is given by
\be
\tilde{h} = e^{i\beta_a (g,v) \tilde{M}^a},
\ee
while the $R$-transformation properties have been suppressed.  The covariant
derivative for the matter field is defined using the Maurer-Cartan spin
connection one-form \cite{Clark:2002bh} 
$\omega_{M}^a =\left( \cosh{\sqrt{v^2}} -1 \right)
\epsilon^{abc} \frac{v_b dv_c}{v^2}$ as
\be
\nabla M \equiv ( d +i \omega_{M}^a \tilde{M}_a ) M .
\ee
It has the same transformation properties as the matter field itself,
\be
(\nabla M)^\prime (x^\prime) = \tilde{h} \nabla M(x) .
\label{Mprime}
\ee
The spin connection one-form, $\omega_{M}^a$, can be expanded in terms of
the covariant coordinate differential basis, $\omega^a$, $\omega_M^a =\omega^a
\Gamma_a^{~b}(= dx^m \Gamma_m^{~b})$ with components
\be
\Gamma_a^{~b} = \left( \cosh{\sqrt{v^2}} -1 \right) \epsilon^{bcd}
\frac{v_c
{\cal D}_a v_d}{v^2} ,
\ee
where ${\cal D}_a = e_a^{-1m} \partial_m$.
As well, expanding the covariant derivative one-form in terms of 
$\omega^a$, the component form of the covariant derivative is obtained
\be
\nabla_a M = \left( {\cal D}_a +i \Gamma_a^{~b} \tilde{M}_b \right) M .
\ee
The component form of the covariant derivative has the $G$ transformation law
\be
\left( \nabla_a M\right)^\prime (x^\prime) = \tilde{h} L_a^{-1 b}
\nabla_b M
(x) .
\ee

The definition of the covariant derivative must be extended when the
matter fields also belong to representations of a local internal symmetry group
${\cal G}$
\be
M^{\prime \alpha}(x) = (U(\epsilon))^\alpha_{~\beta} M^\beta (x) ,
\ee
where the representation matrix
\be
(U(\epsilon))^\alpha_{~\beta} = (e^{ig\epsilon^A (x) T^A})^\alpha_{~\beta},
\ee
is given in terms of the world volume local transformation parameters
$\epsilon^A (x)$, the gauge coupling constant $g$ and the representation generator
matrices $(T^A)^\alpha_{~\beta}$, $A= 1,2,\ldots, {\rm dim}{\left[\cal G\right]}$.
The generators obey the associated Lie algebra $[T^A , T^B ] = if^{ABC} T^C$,
and are normalized so that ${\rm Tr}{[T^A T^B]} = 1/2 \delta^{AB}$.
In order to extend the invariance of the action to include gauge transformations,
the world volume Yang-Mills gauge potential one-form, $A(x)$, must be
introduced, where
\bea
A(x) &=& dx^m A_m (x) \cr
 &=& dx^m (iT^A A^A_m (x)) .
\eea
Under $G$-transformations the Yang-Mills one-form is invariant: $A^\prime
(x^\prime) = A(x)$, that is the world index gauge field transforms as a
coordinate differential
\be
A_m^\prime (x^\prime) = G_m^{-1 n} A_n (x) .
\ee
Under ${\cal G}$-transformations the Yang-Mills field transforms as a
gauge connection
\be
A^\prime = U(\epsilon)AU^{-1}(\epsilon) +\frac{1}{g}
(dU(\epsilon))U^{-1}(\epsilon) .
\ee
Hence, the gauge and super-Poincar\'e covariant derivative of the matter
field is obtained
\be
\nabla M = [d +i\omega^a_M \tilde{M}_{Ma} -gA ] M.
\ee
Under super-Poincar\'e transformations the covariant derivative transforms
as does $M$,
\be
(\nabla M)^\prime (x^\prime) = \tilde{h} (\nabla M)(x) .
\ee
Likewise, under gauge transformations the covariant derivative remains in
the same matter field representation of ${\cal G}$
\be
(\nabla M)^\prime = U(\epsilon) (\nabla M) .
\ee

The matter field derivative can be expanded in terms of the world volume
coordinate differentials $dx^m$ as
\be
\nabla_m M = \left( \partial_m +ie_m^{~a}\Gamma_a^{~b} \tilde{M}_{b} -g
A_m
\right)M .
\ee
The fully covariant derivatives for the scalar, $S(x)$, and fermion,
$\psi_i(x)$, matter fields have the explicit form
\bea
(\nabla_m S)^\alpha &=& \partial_m S^\alpha -igA_m^A (T^A)^\alpha_{~\beta} S^\beta \cr
(\nabla_m \psi)^\alpha_i &=& \partial_m \psi_i^\alpha -\frac{i}{2}
\Gamma_m^{~a}
(\gamma_a)_{ij} \psi_j^\alpha -ig A_m^A
(T^A)^\alpha_{~\beta} \psi^\beta_i  .
\eea
Employing the world volume induced metric, $g^{mn}$, and dreibein,
$e_a^{-1 m}$, the matter field fully invariant action is given by
\be
\Gamma_{\rm matter} = \int d^3 x \det{e}~{\cal L}_{\rm matter},
\ee
where the fully invariant matter field Lagrangian ${\cal L}_{\rm matter}$,
which takes
the form
\bea
{\cal L}_{\rm matter} &=& {\rm Tr}\left[ (\nabla_m S)^\dagger g^{mn}
(\nabla_n
S)\right] - V(S) \cr
 & &\qquad +i\bar\psi \gamma^a e_a^{-1 m} \nabla_m \psi - \bar\psi m \psi +
Y(S,\bar\psi \psi),
\label{matlag}
\eea
includes a globally ${\cal G}$-invariant scalar field potential $V(S)$, and
(if possible to form) globally ${\cal G}$-invariant
fermion mass terms $\bar\psi m \psi$, and generalized Yukawa couplings $Y(S,\bar\psi\psi)$.

The Yang-Mills field strength two-form, $F$, is defined as
\be
F\equiv dA + g A\wedge A .
\ee
As a two-form, $F$ is invariant under $G$-transformations while under 
${\cal G}$-transformations it is in the adjoint representation
\be
F^\prime = U(\epsilon) F U^{-1}(\epsilon) .
\ee
Expanding $F$ in terms of the coordinate differential basis $dx^m$,
$F=\frac{1}{2} dx^n \wedge dx^m (iT^A F^A_{mn})$, the world index field
strength tensor is obtained
\be
F^A_{mn} = \partial_m A_n^A -\partial_n A^A_m + g f^{ABC} A_m^B A_n^C .
\ee
The fully invariant Yang-Mills action (constructed in
\cite{Clark:1997aa} when only supersymmetry is realized
nonlinearly)
is secured as
\be
\Gamma_{\rm Y-M} = \int d^3 x \det{e}~{\cal L}_{\rm Y-M} ,
\ee
with the invariant Lagrangian ${\cal L}_{\rm Y-M}$ given by
\be
{\cal L}_{\rm Y-M} = -\frac{1}{2} {\rm Tr}{[F_{mn} g^{mr} g^{ns} F_{rs}]}.
\label{ymlag}
\ee

There are two other useful one-form bases in which to express the
derivatives and gauge fields.  The basis consisting of the fully covariant world
volume coordinate differentials $\omega^a = dx^m e_m^{~a}$ and the basis
consisting of the partially (SO(1,2)) covariant Akulov-Volkov one-forms 
$\hat\omega^a$ with $\omega^a = \hat{\omega}^b N_b^{~a}$.  
The exterior derivative can be expanded in these bases as 
$d=dx^m \partial_m = \omega^a {\cal D}_a =\hat{\omega}^a\hat{\cal D}_a$ 
where the derivatives are related by the Nambu-Goto dreibein, 
${\cal D}_a = N_a^{-1 b} \hat{\cal D}_b$. Likewise, the gauge field
one-form has the expansion
\be
A = dx^m A_m = \omega^a A_a = \hat{\omega}^a \hat{A}_a .
\ee
As previously noted, the fully covariant basis $\omega^a$ transforms
according to the vector representation of the D=3 (local) Lorentz structure group,
$\omega^{\prime~a} = \omega^b L_b^{~a}$ 
with the Nambu-Goldstone field dependent matrix $L_b^{~a}$ given in 
equations (\ref{L}) and (\ref{beta}).
So the fully covariant gauge
field transforms analogously, $A^\prime_a (x^\prime) = L_a^{-1 b} A_b (x)$.
The
partially covariant basis differentials $\hat{\omega}^a$ transform SO(1,2)
covariantly but not SO(1,3) covariantly,
\bea
\hat{\omega}^{\prime a} &=& \hat{\omega}^b \hat{L}_b^{~a} \cr
&=& \hat{\omega}^b \left( \delta_b^{~a} + \alpha^c \epsilon_{cb}^{~~~a} +
v_b b^a \frac{\tanh{\sqrt{v^2}}}{\sqrt{v^2}} \right) ,
\eea
with $v^a$ related to the Nambu-Goldstone fields via the
\lq\lq inverse Higgs" mechanism, equation (\ref{inversehiggs}),
and likewise for the partially covariant gauge fields
\be
\hat{A}_a^\prime (x^\prime) = \hat{L}_a^{-1 b} \hat{A}_b (x) .
\ee
In similar fashion, the fully covariant derivative transforms as ${\cal
D}_a^\prime = L_a^{-1 b} {\cal D}_b$ while the SO(1,2) partially covariant
Akulov-Volkov derivative transforms as $\hat{\cal D}_a^\prime =
\hat{L}_a^{-1 b}\hat{\cal D}_b$. The invariant interval can be expressed 
in each basis by use of a metric specific to it
\be
ds^2 = dx^m g_{mn} dx^n = \omega^a \eta_{ab} \omega^b = \hat{\omega}^a n_{ab}
\hat{\omega}^b ,
\ee
with $\eta_{ab}$ the flat tangent space Minkowski metric and $n_{ab}$ the
Nambu-Goto metric made from the Nambu-Goto dreibein $N_a^{~b}$, $n_{ab} =
N_a^{~c}\eta_{cd} N_b^{~d} = \eta_{ab} - \frac{v_a v_b}{v^2} \tanh^2{\sqrt{v^2}}$.

The matter field covariant derivatives can be expressed in each basis as
\bea
(\nabla_a S)^\alpha &=& {\cal D}_a S^\alpha -igA_a^A (T^A)^\alpha_{~\beta} S^\beta \cr
(\hat{\nabla}_a S)^\alpha &=& \hat{\cal D}_a S^\alpha -ig\hat{A}_a^A 
(T^A)^\alpha_{~\beta} S^\beta \cr
(\nabla_a \psi )^\alpha_i &=& {\cal D}_a \psi^\alpha_i
-\frac{i}{2}\Gamma_a^{~b}
(\gamma_b)_{ij} \psi^\alpha_j -igA_a^A (T^A)^\alpha_{~\beta} \psi_i^\beta \cr
(\hat{\nabla}_a \psi )^\alpha_i &=& \hat{\cal D}_a \psi^\alpha_i -
\frac{i}{2}\hat{\Gamma}_a^{~b} (\gamma_b)_{ij} \psi^\alpha_j
-ig\hat{A}_a^A (T^A)^\alpha_{~\beta} \psi_i^\beta ,
\eea
with $\hat{\Gamma}_a^{~b} = N_a^{~c} \Gamma_c^{~b}$.  With these
replacements, the fully invariant matter field Lagrangian, equation 
(\ref{matlag}), can be written in the two bases as
\bea
{\cal L}_{\rm matter} &=& {\rm Tr}\left[ (\nabla_a S)^\dagger \eta^{ab}
(\nabla_b S)\right] - V(S) \cr
 & &\qquad +i\bar\psi \gamma^a  \nabla_a \psi - \bar\psi m \psi +
Y(S,\bar\psi \psi) \cr
{\cal L}_{\rm matter} &=& {\rm Tr}\left[ (\hat{\nabla}_a S)^\dagger n^{ab}
(\hat{\nabla}_b S)\right] - V(S) \cr
 & &\qquad +i\bar\psi \hat{\gamma}^a  \hat{\nabla}_a \psi - \bar\psi m \psi
+ Y(S,\bar\psi \psi) ,
\eea
with the Dirac matrices in the partially covariant basis defined by means
of the Nambu-Goto dreibein $\hat{\gamma}^a = \gamma^b N_b^{-1 a}$.

The Yang-Mills fields expressed in terms of the different bases have
correspondingly modified gauge variations
\bea
A^\prime_a &=& U(\epsilon)A_a U^{-1}(\epsilon) +\frac{1}{g} ({\cal D}_a
U(\epsilon))U^{-1}(\epsilon) \cr
\hat{A}^\prime_a &=& U(\epsilon)\hat{A}_a U^{-1}(\epsilon) +\frac{1}{g}
(\hat{\cal D}_a U(\epsilon))U^{-1}(\epsilon) .
\eea
Also, the field strength tensor takes the various forms
\bea
F^A_{ab} &=& {\cal D}_a A_b^A - {\cal D}_b A^A_a + g f^{ABC} A_a^B A_b^C
\cr
\hat{F}^A_{ab} &=& \hat{\cal D}_a \hat{A}_b^A - \hat{\cal D}_b \hat{A}^A_a
+ g f^{ABC} \hat{A}_a^B \hat{A}_b^C .
\eea
As with the matter field Lagrangian, the fully invariant Yang-Mills
Lagrangian, equation (\ref{ymlag}), in these bases becomes
\bea
{\cal L}_{\rm Y-M} &=& -\frac{1}{2} {\rm Tr}{[F_{ab} \eta^{ac} \eta^{bd}
F_{cd}]} \cr
{\cal L}_{\rm Y-M} &=& -\frac{1}{2} {\rm Tr}{[\hat{F}_{ab} n^{ac} n^{bd}
\hat{F}_{cd}]}.
\eea
\vspace{0.25in}

The work of TEC and MN was supported in part by the 
U.S. Department of Energy under grant DE-FG02-91ER40681 (Task B). 

\newpage
\end{document}